\documentstyle[12pt,
prd,epsf,aps,amsmath,amssymb,graphicx,preprint
]{revtex}  
\draft
\begin{document} 
\preprint{DPNU-01-04}
\title{Asymptotic  tails of massive scalar fields
in Schwarzschild background}
\author{Hiroko Koyama\thanks{Email: hiroko@allegro.phys.nagoya-u.ac.jp}
 and Akira Tomimatsu\thanks{Email: atomi@allegro.phys.nagoya-u.ac.jp}}
\address{Department of Physics, Nagoya University, Nagoya 464-8602, Japan}
\date{\today}
\maketitle  
%%%%%%%%%%%%%%%%%%%%%  abstract  %%%%%%%%%%%%%%%%%%%%%%%%%%%%%%%%%%%%%%%%%%%%
\begin{abstract}
We investigate the asymptotic tail behavior of massive scalar
fields in Schwarzschild background.
It is shown that the oscillatory tail of the scalar field
has the decay rate
of $t^{-5/6}$ at asymptotically late times, and 
the oscillation with
the period  $2\pi/m$ 
for the field mass $m$
is modulated by 
the long-term phase shift.
These behaviors are qualitatively similar to those found
in nearly extreme Reissner-Nordstr\"{o}m background,
which are  discussed in terms of a resonant backscattering due to
the space-time curvature.
\end{abstract}
%%%%%%%%%%%% introduction %%%%%%%%%%%%%
\section{Introduction} 
The late-time evolution of various fields outside a collapsing star
has  important implications for several aspects of black-hole physics.
For example, 
the no-hair theorem \cite{W} introduced by Wheeler in the early 1970's,
states that the external field of a black hole relaxes to a Kerr-Newman field
characterized solely by the black-hole's mass, charge and angular-momentum.
Thus, it is of interest to explore the dynamical mechanism
responsible for the relaxation of perturbations fields outside a
black hole and to determine the decay rate of the various
perturbations.
In addition, the dynamical mechanism of generating perturbation
fields ingoing to a black hole is of interest in relation to the problem of
stability of the Cauchy horizon \cite{PandI}.

It was first demonstrated by Price \cite{Price},
regarding scalar, gravitational and electromagnetic perturbations of
Schwarzschild black hole exterior, that the fields die off at late
time with an inverse power-law tail. 
It has been shown that for  a spherical-harmonic wave
mode of multiple number $l$, 
a $t^{-2l -2}$ or $t^{-2l-3}$
decay tail ($t$ being the Schwarzschild time coordinate) 
dominates at late time,
depending on the initial conditions. 
The same tail behavior occurs also 
for Reissner-Nordstr\"{o}m black holes and fields of different spin.
Recently the linear perturbation analyses and the nonlinear simulations 
have been done by Gundlach et al. \cite{GPP1} and \cite{GPP2}, respectively.
(See also \cite{BuO}.)
In addition, the recent treatment of
tails in the spacetime of a spinning black hole has been 
done by Barack and Ori \cite{BaO}.

%%%%%%%%%% K-K %%%%%%%%%%%%%%%%
Many previous works are mainly concerned with massless fields.
However, the evolution of massive scalar fields will become important, 
for example,
if one considers Kaluza-Klein theories, in which the Fourier modes 
can behave like massive fields.  
Further, the 
recent development of the Kaluza-Klein idea, 
such as Rendall-Sundrum model \cite{R-Smodel} in string theory,
strongly motivates us to understand 
the evolutional features characteristic to the field mass in detail.

%%%%%%%% massive tail intermediate %%%%%%%%%%
The physical mechanism by which 
late-time tails of massive scalar
fields %in nearly extreme Reissner-Nordstr\"{o}m background
are generated 
may be qualitatively different from that of  
massless ones. 
In fact, it has been pointed out that
the oscillatory inverse power-law tails
\begin{equation}
  \label{eq:HP}
  \psi \sim t^{\scriptscriptstyle - l-\frac 32}\sin(mt),
\end{equation}
dominates as the intermediate late-time, 
if the field mass $m$ is small
\cite{HandP} (see also \cite{Burko}).
It is clear from Eq. (\ref{eq:HP}) that 
massive fields decay slower than massless ones, and 
waves with peculiar frequency $\omega$ quite close to $m$
can  contribute to massive tail,
while low frequency waves can contribute to massless tail.
Though the oscillatory power-law form (\ref{eq:HP}) 
has been numerically verified at
intermediate late times, $mM \ll mt\ll 1/(mM)^2$,
it should be  noted that the intermediate tails are
not the final asymptotic behaviors;
Another wave pattern can dominate at very late times, when 
it still remains  very difficult to determine numerically 
the exact decay rate \cite{HandP,Burko}.

%%%%%%%% massive tail 2 %%%%%%%%%%
In the previous paper \cite{KandT},
hereafter referred to as paper I,
we have analytically found  that 
the transition from the intermediate behavior to the asymptotic  one
occurs
in nearly extreme Reissner-Nordstr\"{o}m background.
The oscillatory inverse power-law behavior of 
the dominant asymptotic tail is approximately given by 
\begin{eqnarray}
  \label{eq:asym_tail}
  \psi \sim t^{-5/6}\sin (mt),
\end{eqnarray}
independently of angular momentum wave number $l$, and the 
decay becomes  slower than the intermediate ones.
The origin of such an 
asymptotic tail is expected to be a resonance backscattering
due to curvature-induced potential.
%In addition, we have estimated
This will be supported by
the relationship between 
the field mass and
the time scale when the $t^{-5/6}$ tail dominates;
For $mM \ll 1$, 
the smaller $mM$ is,
the later the $t^{-5/6}$ tail begins to dominate 
($mt \gg 1/(mM)^2$, where $M$ is the black-hole mass).
On the other hand, 
for large field mass $mM \gg 1$,
the larger $mM$ is,
the later the $t^{-5/6}$ tail begins to dominate ($mt \gg mM$).
Therefore, the time scale
when the $t^{-5/6}$ tail dominates
will become minimum
%soonest 
at $mM \simeq O(1)$,
%In other words, 
which means that the most effective backscattering occurs 
for such massive scalar fields 
with the Compton wave length $1/m$ nearly equal to the horizon 
radius $M$.

In this paper, following paper I which was devoted 
to the analysis for nearly extreme Reissner-Nordstr\"{o}m background,
we investigate the 
asymptotic tails of a massive scalar field
in Schwarzschild background.
The purpose of this paper is 
to determine analytically the decay rate and to confirm that 
the asymptotic tail with the  decay rate of $t^{-5/6}$ is not 
peculiar to the nearly extreme black hole.
Nevertheless,
it is sure that
 the form of effective potential varies  according to the ratio
of the black-hole charge to the mass.
Then, we also study the difference of the 
asymptotic tail behavior in  Schwarzschild background 
from that in nearly extreme Reissner-Nordstr\"{o}m.
In Sec. II, to analyze time evolution of massive scalar fields,
we introduce the black-hole Green's function using the spectral
decomposition method \cite{Leaver}.
Differently from the case of nearly extreme Reissner-Nordstr\"{o}m
in paper I,
it is difficult to analyze the field equations for the  
arbitrary field mass $m$.
In this paper, therefore, we consider the cases of 
small field mass $mM \ll 1$ in Sec. III 
 and large one $mM \gg 1$ in Sec. IV,
%where $M$ is the black-hole mass.
%In Sec. III and IV we study the asymptotic tail in the small mass case.
%and in the large mass case, respectively, 
and we compare the results with those found 
in nearly extreme Reissner-Nordstr\"{o}m background.
Section V is devoted to a summary.

\section{Green's function analysis}
\subsection{Massive scalar fields in Schwarzschild geometry}
We consider time evolution of a massive scalar field in Schwarzschild 
background with the black-hole mass $M$. The metric is
\begin{equation}
  ds^2=-\left(1-\frac{2M}{r}\right)dt^2
+\left(1-\frac{2M}{r}\right)^{-1}dr^2
+r^2d\Omega ^2,
\end{equation}  
and the scalar field $\phi$ with the mass $m$ satisfies the wave equation
\begin{equation}
  \label{eq:K-G}
  (\Box -m^2)\phi =0.
\end{equation}
Resolving the field into spherical harmonics
\begin{equation}
  \label{eq:harmonics}
  \phi =\sum _{l,m}\frac{\psi ^l(r)}{r}Y_{lm}(\theta ,\varphi),
\end{equation}
hereafter we omit the index $l$ of $\psi ^l$ for simplicity,
and we obtain a wave equation for each multiple moment:
\begin{equation}
  \psi _{,tt}-\psi _{,r_{\ast}r_{\ast}}+V\psi =0,
\end{equation}
where $r_{\ast}$ is the tortoise coordinate defined by
\begin{equation}
  \label{eq:tortoise}
  dr_{\ast}=\frac{dr}{1-\frac{2M}{r}}\>,
\end{equation}
and the effective potential $V$ is
\begin{equation}
  \label{eq:potential}
  V=\left(1-\frac{2M}{r}\right)
\left[\frac{l(l+1)}{r^2}+\frac{2M}{r^3} +m^2\right].
\end{equation}
%%%%%%%%%%%%%%%%%%%%%%%%%%%%%%%%%%%%%%%%%%%%%%%%%%%%%%%%%%%%%%
\subsection{The black-hole Green's function}
\begin{figure}
\begin{center}  
\leavevmode
\epsfxsize=80mm
\epsfbox{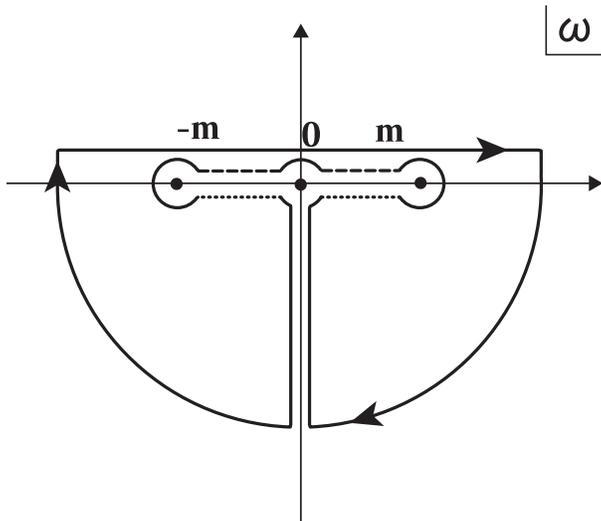}  
\caption{Integration contours in the complex frequency plane.
The original integration contour for the Green's function lies above 
the real frequency axis.
We choose the value of $\varpi$ on the dashed line to be  $\varpi = |\varpi|$
and that on the dotted lines to be  $\varpi =e^{\pm i\pi}|\varpi|$.}
\label{fig1}
\end{center}
\end{figure}

The time evolution of a massive scalar field is given by
\begin{equation}
  \label{eq:evolution}
  \psi (r_{\ast},t)=\int \left[G(r_{\ast},r_{\ast}';t)\psi _t(r',0) 
+ G_t(r_{\ast},r_{\ast}';t)\psi (r_{\ast}',0) \right]
dr_{\ast}'
\end{equation}
for $t>0$, where the (retarded) Green's function $G(r,r';t)$ is defined as
\begin{equation}
  \label{eq:retarded}
  \left[\frac{\partial ^2}{\partial t^2} 
-\frac{\partial ^2}{\partial r_{\ast}^2} 
+V
\right]G(r_{\ast},r_{\ast}';t) = \delta (t)\delta (r_{\ast}-r_{\ast}').
\end{equation}
The causality condition gives the initial condition that 
$G(r,r';t)=0$ for $t\le 0$. In order to find $G(r,r';t)$
we use the Fourier transform
\begin{equation}
  \label{eq:fourier}
  \tilde{G}(r_{\ast},r'_{\ast};\omega)
=\int _{0^{-}}^{+\infty} G(r_{\ast},r'_{\ast};t)e^{i\omega t}dt.
\end{equation}
The Fourier transform is analytic in the upper half $\omega$ plane,
and the corresponding inversion formula is
\begin{equation}
  \label{inverse}
  G(r_{\ast},r_{\ast}';t)= -\frac{1}{2\pi}\int _{-\infty +ic}^{\infty +ic} 
\tilde{G}(r_{\ast},r'_{\ast};\omega)
e^{\scriptscriptstyle -i\omega t}d\omega
\end{equation}
where $c$ is some positive constant.
The usual procedure is to
close the contour of integration into the lower half of the 
complex frequency plane shown in Fig. \ref{fig1}. 
Then, the late-time tail behaviors which are our main concern should be
given by the integral along the branch cut
placed along the interval $-m \le \omega \le m$.
%%%%%%%%%%%%%%%%%%%% figure %%%%%%%%%%%%%%%%%%%%%%%%%%%%%%%%%%%%%%%%%%%%%%%%%%

Now the Fourier component of 
the Green's function $\tilde{G}(r_{\ast},r'_{\ast};\omega)$
 can be expressed in terms of two linearly independent solutions
for the homogeneous equation
\begin{equation}
  \label{eq:homo}
  \left(\frac{d^2}{dr_{\ast}^2}+\omega ^2 -V\right)
\tilde{\psi}_i =0 \quad i=1,2.
\end{equation}
The boundary condition for the basic solution $\tilde{\psi}_1$
is that it describes purely ingoing waves crossing the event horizon, i.e.,
\begin{equation}
  \label{eq:boundary1}
  \tilde{\psi}_1 \simeq e^{-i\omega r_{\ast}}, 
\end{equation}
at $r_{\ast} \to -\infty$,
while $\tilde{\psi}_2$ is required to damp exponentially
at spatial infinity, i.e.,
\begin{equation}
  \label{eq:boundary2}
  \tilde{\psi}_2 \simeq  e^{-\varpi r_{\ast}},
\end{equation}
at $r_{\ast} \to \infty$,
where $\varpi \equiv \sqrt{m^2-\omega ^2}$.
Because the complex conjugate $\tilde{\psi}_1^{\ast}$ is also a solution for 
Eq. (\ref{eq:homo}),
$\tilde{\psi}_2$ can be written by the linear superposition
\begin{equation}
  \label{eq:2to1}
  \tilde{\psi}_2 =  
\alpha \tilde{\psi}_1 +\beta\tilde{\psi}_1^{\ast},
\end{equation}
and the Wronskian is estimated to be
\begin{equation}
  \label{}
W(\omega )=\tilde{\psi}_1\tilde{\psi}_{2,r_{\ast}}
-\tilde{\psi}_{1,r_{\ast}}\tilde{\psi}_2 =2i\omega \beta.
\end{equation}
%%%%%%%%%%%%%%%%%%%%%%%%%%%%%%%%%%%%%%%%%%%%%%%%%%%%%%%%%%%%%%%%
Using these two solutions, the Green's function can be written by
\begin{equation}
  \label{eq:Green}
 \tilde{G}(r_{\ast},r'_{\ast};\omega )= -\frac{1}{2i\omega \beta}
\left\{ 
\begin{array}{l@{\quad,\quad}l}
\tilde{\psi} _1(r'_{\ast},\omega)
\tilde{\psi }_2(r_{\ast},\omega)&
\qquad  r'_{\ast} >r _{\ast},\\
\tilde{\psi} _1(r_{\ast},\omega)
\tilde{\psi }_2(r'_{\ast},\omega)
&\qquad r'_{\ast}< r_{\ast} .
\end{array}
\right.
\end{equation}  
The contribution $G^C$ from the branch cut to the Green's function
is reduced to 
\begin{eqnarray}
  \label{eq:branch-cut2}
  G^C (r_{\ast},r_{\ast}';t)&=&
-\frac 1{4\pi i}
\int _{\rm{cut}}
\frac 1\omega
\frac{\alpha}{\beta}
\tilde{\psi}_1(r_{\ast}',\omega)
\tilde{\psi}_1(r_{\ast},\omega)
e^{-i\omega t}d\omega .
\end{eqnarray}
Then the main task to evaluate $G^C$ is to derive the coefficients
$\alpha$ and $\beta$.
%%%%%%%%%%%%%%%%%%%%%%%%%%%%%%%%%%%%%%%%%%%%%%%%%%%%%%%%%%%
\subsection{Effective contributions to tail behaviors}
Here we point out the at very late times $mt\gg 1$
 the rapidly oscillating term $e^{-i\omega t}$ leads to a mutual 
cancellation between the positive and the negative parts 
of the integrand in Eq.(\ref{eq:branch-cut2}).
Then, the effective contribution to the tail behaviors arises from the narrow 
range
\begin{eqnarray}
\label{near_bound}
\epsilon \equiv 2M\varpi \ll 1,
\end{eqnarray}
which was also claimed in \cite{HandP,TandK}.

%%%%%%%%%%%%%%%%%%%%  small mass  case  %%%%%%%%%%%%%%%%%%%% 
\section{small-mass  case}
Introducing the non-dimensional variable defined as
\begin{eqnarray}
  x&\equiv&\frac {r}{2M},
\end{eqnarray}
Eq.(\ref{eq:homo}) is rewritten by
\begin{eqnarray}
\label{eq:homo2}
&&  \frac{d^2\tilde{\psi}}{dx^2}
+\frac{1}{x(x-1)}\frac{d\tilde{\psi}}{dx}
+\Bigg[\frac{4M^2\omega ^2x^2}{(x-1)^2}
-\frac{4M^2m^2x}{x-1}
-\frac{l(l+1)}{x(x-1)}
-\frac{1}{x^2(x-1)}
\Bigg]\tilde{\psi} =0.
\end{eqnarray}
To study analytically the mode solutions $\psi$,
the asymptotic matching between the inner and outer solutions was 
successfully used in paper I for the potential $V$ 
in nearly extreme Reissner-Nordstr\"{o}m background.
However, to apply the same method to Eq.(\ref{eq:homo2}),
it is necessary to assume the field mass to be very small or very large.
In this section we consider the small-mass case $mM \ll 1$.

\subsection{Mode solutions}
\subsubsection{The inner region $1 \le x\ll 1/m^2M^2$}
For the small-mass case such as $ mM \ll 1$,
truncating the terms 
of the  order $m^2M^2$,
we can approximate Eq.(\ref{eq:homo2}) by
\begin{eqnarray}
\label{eq:homo3}
   \frac{d^2\tilde{\psi}}{dx^2}
+\left(\frac{1}{x-1}-\frac{1}{x}\right)\frac{d\tilde{\psi}}{dx}
+\Bigg[\frac{4M^2m^2}{(x-1)^2}
+\frac{4M^2m^2-l^2-l-1}{x(x-1)}
+\frac{1}{x^2}
\Bigg]\tilde{\psi} &=&0.
\end{eqnarray}
Now the solution $\psi _1$ satisfying the boundary condition 
(\ref{eq:boundary1})
can be written using the hyper-geometric function $F$ as follows,
\begin{eqnarray}
  \label{eq:PIM}
\psi _1&=& 
x^{K_{++}}(x-1)^{-2iMm}
F\left(K_{--},K_{--}  ; 
1-4iMm ;
1-\frac 1x\right)
\nonumber\\
&=&
x^{K_{++}}(x-1)^{-2iMm}
\frac{\Gamma(1-4iMm)\Gamma(2\mu)}
{\Gamma(K_{+-})^2}
F\left(K_{--},K_{--}; 
1-2\mu ;\frac 1x\right)
\nonumber\\&&
+
x^{K_{-+}}(x-1)^{-2iMm}
\frac{\Gamma(1-4iMm)\Gamma(-2\mu)}
{\Gamma(K_{--})^2}
F\left(K_{+-},K_{+-} ; 
1+2\mu ;\frac 1x\right),
\end{eqnarray}
where $\mu$ and $K_{\pm\pm}$ are
\begin{eqnarray}
  \mu &=& \sqrt{\left(l+\frac 12\right)^2-8M^2m^2}
\end{eqnarray}
and
\begin{eqnarray}
  K_{\pm\pm} &=& \frac 12 \pm \mu \pm 2iMm,
\end{eqnarray}
respectively.
and we used the linear transformation formulas (15.3.6)
of \cite{Abramo}
in the second equality of 
Eq. (\ref{eq:PIM}).
If estimated in the region $x\gg 1$, we obtain
\begin{eqnarray}
\label{asy_psi1}
  \tilde{\psi} _1
&\to&
\frac{\Gamma(1-4iMm)\Gamma(2\mu)}
{\Gamma(K_{+-})^2}
x^{\mu +\frac 12}
+
\frac{\Gamma(1-4iMm)\Gamma(-2\mu)}
{\Gamma(K_{--})^2}
x^{-\mu +\frac 12},
\end{eqnarray}
which is necessary for asymptotic matching in the overlap region
$1 \ll x \ll 1/(mM)^2$
with the outer solutions valid in the region $x\gg 1$.
\subsubsection{The outer region $x \gg 1$}
For large $x$, Eq.(\ref{eq:homo2}) is approximated by
\begin{eqnarray}
\label{eq_psi2}
    \frac{d^2\tilde{\psi}}{dx^2}
+\Bigg[\frac{4M^2m^2}{x}
-\epsilon ^2
+\frac{8M^2m^2-l(l+1)}{x^2}
\Bigg]\tilde{\psi} &=&0.
\end{eqnarray}
Then we can write the solutions using Whittaker's functions.
The mode solution $\psi _2$ satisfying the boundary condition 
(\ref{eq:boundary2}) should be
\begin{eqnarray}
  \psi _2
&=&W_{\sigma,\mu}(2\epsilon x),
\end{eqnarray}
where
\begin{eqnarray}
  \sigma &=& \frac{2M^2m^2}{\epsilon}-\epsilon .
\end{eqnarray}
If estimated in the region $x\ll 1/\epsilon$, we obtain
\begin{eqnarray}
\label{asy_psi2}
  \tilde{\psi} _2
&\to&
\frac{\Gamma(-2\mu)}{\Gamma(\frac 12 -\mu -\sigma)}
(2\epsilon)^{\mu+\frac 12}
x^{\mu+\frac 12}
+\frac{\Gamma(2\mu)}{\Gamma(\frac 12 +\mu -\sigma)}
(2\epsilon)^{-\mu+\frac 12}
x^{-\mu+\frac 12}.
\end{eqnarray}

\subsubsection{Matching}
We can match the solution (\ref{asy_psi1}) valid
at $1\ll x\ll 1/m^2M^2$ with the solution
(\ref{asy_psi2}) valid at $1\ll x \ll 1/\epsilon $
in the overlap region
and determine the coefficients $\alpha$ and $\beta$ 
in Eq.(\ref{eq:branch-cut2}).
Then the ratio is given as follows,
\begin{eqnarray}
  \frac{\alpha}{\beta}(|\omega|,\epsilon)\Bigg|_{|\omega|\to m}
&=&\Bigg[
-(2\epsilon)^{\mu +\frac 12}
\frac{\Gamma(-2\mu)^2\Gamma(1+4iMm)}
{\Gamma(\frac 12 -\mu -\sigma)
%\Gamma(\frac 12 -\mu +2iMm)^2}
\Gamma(K_{-+})^2}
\nonumber\\&&
+(2\epsilon)^{-\mu +\frac 12}
\frac{\Gamma(2\mu)^2\Gamma(1+4iMm)}
{\Gamma(\frac 12 +\mu -\sigma)
\Gamma(K_{++})^2}
\Bigg]
\times ({\rm complex\> conjugate})^{-1}.
\end{eqnarray}

%%%%%%%%%%%%%% Intermediate tails %%%%%%%%%%%%%%%%%%%%
\subsection{Intermediate tails}
%Successively \cite{HandP,KandT},
The effective contribution to the integral in Eq. (\ref{eq:branch-cut2})
is claimed to be limited to the range $|\omega - m| =O(1/t)$
%arises from $|\omega | =O(m-1/t)$ 
or equivalently 
$\varpi = O(\sqrt{m/t})$,
since the rapidly oscillating term $e^{-i\omega t}$
which leads to a mutual cancellation between the positive and the negative 
parts of the integrand (see \cite{HandP,KandT}).
The intermediate tails becomes dominant 
at the  late times in the range
\begin{eqnarray}
  \label{eq:time_window}
  mM \ll mt \ll \frac{1}{(mM)^2},
\end{eqnarray}
when the integral (\ref{eq:branch-cut2}) 
should be estimated under the condition
\begin{eqnarray}
\label{nonback_time}
  \sigma \simeq \frac{2m^2M^2}{\epsilon} 
= O(mM\sqrt{mt}) \ll 1.
\end{eqnarray}
As was discussed in paper I,
the small value of $\sigma$ represents that the backscattering due to 
the spacetime curvature is not effective at intermediate late times.
%a degree of the domination of the backscattering.
%Just like paper I, 
%approximating $\alpha/\beta$ 
Then, using the condition (\ref{nonback_time}),  
we obtain
\begin{eqnarray}
\frac{\alpha}{\beta}(|\omega|,\epsilon)
-\frac{\alpha}{\beta}(|\omega|,e^{-i\pi}\epsilon)
\Bigg|_{|\omega|\to m}
\sim
  i\frac{2^{2l+1}l! ^6}{(2l)!^2(2l+1)!^4}
mM
\epsilon ^{2l+1},
\end{eqnarray}
which is identical with Eq. (51) in paper I.
%Substituting it into Eq.(\ref{eq:branch-cut2}), 
Therefore, it is obvious that the same
 intermediate tails as Eq. (\ref{eq:HP}) dominate
 at intermediate late times (\ref{eq:time_window}),
which was numerically supported by \cite{HandP,Burko}.
%%%%%%%%%%%% Asymptotic tails %%%%%%%%%%%%%%%%%%
\subsection{Asymptotic tails}
As discussed in paper I,
the intermediate tails 
%dominate at the intermediate late times 
%(\ref{eq:time_window}), which 
can not be an asymptotic behavior,
and the long-term evolution 
from the intermediate behavior to the final one should occur.
The asymptotic tail becomes dominant at very late times such that 
\begin{eqnarray}
  mt \gg \frac{1}{m^2M^2},
\end{eqnarray}
when the effective contribution to the integral (\ref{eq:branch-cut2}) arises 
from the region
\begin{eqnarray}
\label{asympto_time}
  \sigma \simeq \frac{2m^2M^2}{\epsilon} \gg 1,
\end{eqnarray}
which means backscattering effect due to curvature-induced potential
dominates. 
%at this stage.
Then, the ratio of $\alpha$ to $\beta$ is approximately given by
\begin{eqnarray}
   \frac{\alpha}{\beta}
(|\omega|,\epsilon)
\Bigg|_{|\omega|\to m}
&\to&
\frac{\eta ^{\ast} e^{i\pi\sigma}+\gamma^{\ast} e^{-i\pi\sigma}}
{\eta e^{-i\pi\sigma}+\gamma e^{i\pi\sigma}},
\end{eqnarray}
where
\begin{eqnarray}
\eta&=&
-\frac{\Gamma(-2\mu)^2\Gamma(1-4iMm)}
{\Gamma(K_{--})^2}
(4m^2M^2)^{\mu}
e^{-i\pi\mu}
+
\frac{\Gamma(2\mu)^2\Gamma(1-4iMm)}
{\Gamma(K_{+-})^2}
(4m^2M^2)^{-\mu}
e^{i\pi\mu},
\end{eqnarray}
and
\begin{eqnarray}
\gamma&=&
-\frac{\Gamma(-2\mu)^2\Gamma(1-4iMm)}
{\Gamma(K_{--})^2}
(4m^2M^2)^{\mu}
e^{i\pi\mu}
+
\frac{\Gamma(2\mu)^2\Gamma(1-4iMm)}
{\Gamma(K_{+-})^2}
(4m^2M^2)^{-\mu}
e^{-i\pi\mu}.
\end{eqnarray}
As was shown in the paper I, 
the contribution from the Green's function to the asymptotic tail part
corresponds to the integral along the dashed line in Fig. \ref{fig1}
which is  approximated by
\begin{eqnarray}
\label{asympto-int}
G^C(r_{\ast},r_{\ast}';t)
&\simeq&
\frac{1}{4\pi mi}  
\tilde{\psi}_1(r_{\ast},m)
\tilde{\psi}_1(r'_{\ast},m)
\int _{\rm dashed\>line}
e^{i(2\pi\sigma -\omega t)} 
e^{i\varphi _{s}}
d\omega +({\rm complex \> conjugate }),
\end{eqnarray}
where the phase $\varphi _{s}$ is defined by
\begin{equation}
\label{phase_small}
  e^{i\varphi _{s}}=\frac{\eta ^{\ast}
+\gamma ^{\ast}e^{-2i\pi \sigma }}
{\eta +\gamma e^{2i\pi \sigma }},
\end{equation}
and it remains in the range $0 \le\varphi _{s}\le 2\pi$, 
even if $\sigma $ becomes very large,
since we have
\begin{eqnarray}
  |\eta|^2-|\gamma|^2
&=&\frac{2\pi Mm}{\mu}\>>\>0.
\end{eqnarray}
Because the terms $e^{-i\omega t}$ and $e^{2i\pi\sigma}$ 
in Eq.(\ref{asympto-int}) are rapidly oscillating at very late times,
the saddle-point integration allows us to evaluate accurately 
the asymptotic behaviors.
Let us introduce the variable $a$ defined by
\begin{eqnarray}
  \label{assume}
a \equiv \left(1-\frac{\omega}{m}\right)^{1/2}(mt)^{1/3}.
\end{eqnarray}
Then, the oscillation terms  $e^{i(2\pi\sigma -\omega t)}$ in 
Eq.(\ref{asympto-int}) can be  approximately rewritten into the form
\begin{eqnarray}
  \label{exp_imt_small}
e^{i(mt)^{1/3}f_{s}(a)}e^{-imt}
\end{eqnarray}
in the limits $mt \to \infty$ and $\omega /m \to 1$, 
by keeping $a$ to be finite. Here we have
\begin{eqnarray}
\label{f(a)_small}
  f_{s}(a)=\frac {\sqrt{2}\pi mM}{a} +a^2.
\end{eqnarray}
The saddle point is found to exist at
\begin{eqnarray}
\label{a_0}
 a= a_0\equiv \left(\frac{\pi mM}{\sqrt{2}}\right)^{1/3}.
\end{eqnarray}
At the saddle point the parameter $\epsilon$ defined 
by Eq.(\ref{near_bound}) is given by
\begin{eqnarray}
  \label{saddle}
\epsilon _0 \equiv  2M\sqrt{m^2-\omega _0^2}
\simeq 2\left(\frac{2\pi }{mt}\right)^{1/3}(mM)^{4/3}, 
\end{eqnarray}
or equivalently
\begin{eqnarray}
  \label{eq:saddle_sigma}
  \sigma _0 
\simeq 
mM\left(\frac{2\pi mM}{mt}\right)^{1/3}
\end{eqnarray}
which is equal to the value obtained in the  nearly extreme case in paper I.
Let us give the Taylor expansion of $f$ near the point $a=a_0$ as follows,
\begin{eqnarray}
  f_{s}(a)&\simeq& f_{s}(a_0)+\frac{1}{2}(a-a_0)^2f_{s}''(a_0)\nonumber\\
&=&3\left(\frac{\pi mM}{\sqrt 2}\right)^{2/3}
+3(a-a_0)^2.
\end{eqnarray}
Then, the integral in Eq.(\ref{asympto-int}) can be 
approximately estimated to be 
\begin{eqnarray}
    \label{eq:massivetail}
 G^{C}(r_{\ast},r'_{\ast};t) 
&\simeq&  \frac{1}{2\sqrt 3}(2\pi)^{\frac 56}
(mM)^{\frac 13}(mt)^{-\frac 56} 
%\sin(\omega _0t -2\pi \sigma (\omega _0)- \varphi _{s}(\omega _0)+3\pi/4 )
\sin(mt -3(2\pi mM)^{2/3}(mt)^{1/3}/2- \varphi _{s}(\sigma _0)+3\pi/4 )
\nonumber\\&&
\times\tilde{\psi}_1(r_{\ast},m)\tilde{\psi}_1(r_{\ast}',m),
\end{eqnarray}
at very late times $mt \gg 1/(mM)^2$.
The oscillation has the period $2\pi /m$,  
and is modulated by
the two types of long-term phase shifts.
First, the term $(3/2)(2\pi mM)^{2/3}(mt)^{1/3} $ represents 
a monotonously increasing phase shift.
Second, the term $\varphi _{s}(\sigma _0)$ represents a periodic phase shift
with the amplitude $|\varphi _{s}|$ smaller than $2\pi$, which will be studied
in the following.
Eq.(\ref{eq:massivetail}) is similar to Eq.(69) in paper I,
though  the corresponding terms of the phase shift are omitted 
in Eq.(69) in paper I
since we have focused our attention to the decay rate.
\subsection{Phase shift}
\begin{figure}
\begin{center}
\leavevmode  
\epsfxsize=80mm              
    \epsfbox{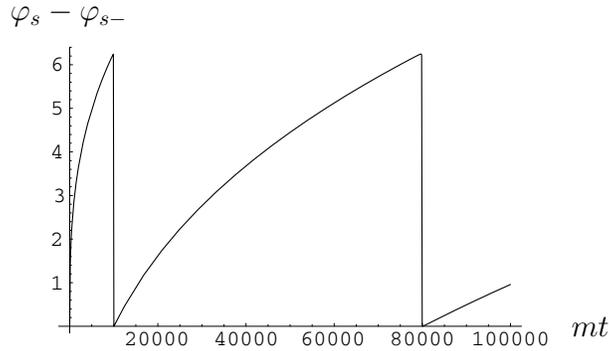}  
\caption{Time evolution of the phase 
$\varphi_s -\varphi_{s-}$ in the case of
$mM =0.01$ and $l=0$. }
\label{small}
\end{center} 
\end{figure}

Now 
we study the periodic phase-shift effect
caused by the term $\varphi _{s}(\omega _0)$.
%The value of which is a function of $\sigma$.
From the  equation
\begin{eqnarray}
  \frac{d \varphi_{s}}{d\sigma}&=&0,
\end{eqnarray}
we find the 
maximum and minimum extremes, denoted by
$\varphi _{s+}(\sigma _+)$ and $\varphi _{s-}(\sigma _-)$ 
of $\varphi_{s}$ at $\sigma =\sigma _+$ and $\sigma =\sigma _-$,
which given by 
\begin{eqnarray}
\label{varphi_spm}
  \varphi _{s\pm}(\sigma _{\pm})&=& \pm 2\theta +\varphi _{s0},
\end{eqnarray}
and 
\begin{eqnarray}
\label{sigma_spm}
  2\pi \sigma _{\pm} &=& \pi \mp\frac{\pi}{2}\pm\theta +const,
\end{eqnarray}
where
\begin{eqnarray}
  \sin \theta &=&\left|\frac{\gamma}{\eta}\right|
\end{eqnarray}
and
\begin{eqnarray}
\label{varphi_s0}
  \exp (i\varphi _{s0})\equiv \frac{\eta ^{\ast}}{\eta}.
\end{eqnarray}
Expanding $|\gamma /\eta|$ with respect to $mM \ll 1$, 
we obtain 
\begin{eqnarray}
  \left|\frac{\gamma}{\eta}\right| &=& 
1-\delta ,
\end{eqnarray}
and
\begin{eqnarray}
  \delta &\simeq& 
\pi
\left(\frac{64}{15}\right)^2
(mM)^3 \qquad ({\rm for} \> l=0),\nonumber\\
&\simeq& 
\frac{2^{4l+4}\pi(l!)^4}{(2l+1)(2l)!^4}
(mM)^{4l+3}
\qquad ({\rm for} \> l\ge0)
\end{eqnarray}
in Schwarzschild background. 
The phase-shift term $\delta$ is characteristic to the black hole.
For example, 
for nearly extreme Reissner-Nordstr\"{o}m background \cite{KandT},
the value of $\delta$ is given by
\begin{eqnarray}
  \delta _{\rm nearly \>extreme }&\simeq&
\pi \left(\frac{50}{23}\right)^2(mM)^3
\qquad ({\rm for} \> l=0),\nonumber\\
&\simeq&
\frac{2^{4l+3}\pi(l!)^2}{(2l+1)^2(2l)!^4}
(mM)^{6l+3}
\qquad ({\rm for} \> l\ge0).
\end{eqnarray}
From Eqs. (\ref{varphi_spm}) and (\ref{sigma_spm}) 
the oscillation amplitude of $\varphi _{s}$
is given by 
\begin{eqnarray}
\label{gap_varphi}
  \Delta \varphi _{s}&\equiv & \varphi _{s+}-\varphi _{s-}
=4\theta \nonumber\\
&\simeq& 2\pi -\sqrt{2\delta},
\end{eqnarray}
and the time interval between the two extremes is estimated to be 
\begin{eqnarray}
\label{gap_sigma}
  2\pi\Delta\sigma 
&\equiv & 2\pi (\sigma _- -\sigma _+)=\pi -2\theta 
\nonumber\\
&\simeq& \sqrt{8\delta}.
\end{eqnarray}
These equations  mean that 
the smaller the field mass is,
the more rapidly 
the change of the phase $\varphi _s$ from the maxixum to the minimum
occurs (see Fig. \ref{small}).
The phase shift oscillates at the time scale of $1/(m^3M^2)$,
which is much longer than the period of $2\pi /m$.
So it is a long-term effect
which modulates the basic oscillation. 
%%%%%%%%%%%%%%%%%%%%%%%  figure %%%%%%%%%%%%%%%%%%%

%%%%%%%%%%%  large mass  %%%%%%%%%%%%%%%%%%%%%%%%%%%%%%
\section{large mass  case}
\subsection{Mode solutions}
In this section we consider large-mass case such as $mM\gg 1$.
It is enough to consider the narrow region (\ref{near_bound})
in the same manner with the small-mass case.
Introducing the function $\xi$ defined as 
\begin{eqnarray}
  \tilde{\psi} =\sqrt{\frac {r}{r-2M}}\xi = \sqrt{\frac {x}{x-1}}\xi ,
\end{eqnarray}
the mode equation (\ref{eq:homo2}) can be approximately given by
\begin{eqnarray}
\label{eq:mode_large}
   \frac{d^2\xi ^2}{dx^2} 
+\Bigg[\frac{4M^2\omega ^2x^2}{(x-1)^2}
-\frac{4M^2m^2x}{x-1}
\Bigg]\xi &=&0,
\end{eqnarray}
since the other terms become negligibly small all over the region $x\ge 1$
in comparison with the two terms
in the potential.
Then we can write solutions for Eq.(\ref{eq:mode_large}) 
using Whittaker's functions.
The solutions  $\psi _1$ and $\psi _2$ 
satisfying the boundary condition 
(\ref{eq:boundary1}) and (\ref{eq:boundary2}), respectively,
are
\begin{eqnarray}
\label{phi1_large}
  \tilde{\psi} _1 &=& \sqrt{\frac{x}{x-1}}M_{\sigma ,\rho}
\left(2\epsilon(x-1)\right),
\end{eqnarray}
and
\begin{eqnarray}
\label{phi2_large}
  \tilde{\psi _2} &=& \sqrt{\frac{x}{x-1}}W_{\sigma ,\rho}
\left(2\epsilon(x-1) \right),
\end{eqnarray}
where
\begin{eqnarray}
\rho &=&-i\sqrt{4M^2\omega ^2-\frac 14}.
\end{eqnarray}
From Eqs. (\ref{phi1_large}) and (\ref{phi2_large}), 
$\psi _2$ can be rewritten as
\begin{equation}
  \tilde{\psi} _2 =
\frac{\Gamma(-2\rho)}{\Gamma(\frac 12 -\rho -\sigma)}\psi _1 
+\frac{\Gamma(2\rho)}{\Gamma(\frac 12 +\rho -\sigma)}\psi _1^{\ast} 
\end{equation}
and the ratio of $\alpha$ to  $\beta$ in Eq.(\ref{eq:2to1}) is
\begin{eqnarray}
  \frac {\alpha}{\beta} &=& 
\frac{\Gamma(-2\rho)\Gamma(\frac 12 +\rho -\sigma)}
{\Gamma(2\rho)\Gamma(\frac 12 -\rho -\sigma)}.
\end{eqnarray}
%%%%%%%%%%%%%%%%%%%%%%%%%%%%%%%
\subsection{Asymptotic tail}
As was discussed in paper I,
no intermediate tails appear 
for  the scalar field with large mass, namely $mM\gg 1$. 
At late times given by Eq.(\ref{asympto_time}),
the ratio of $\alpha$ to  $\beta$ is approximately written
by
\begin{eqnarray}
\label{exp_large-mass}
  \frac {\alpha}{\beta} 
(|\omega|,\epsilon)
\Bigg|_{|\omega|\to m}
&\to&
\frac{\zeta ^{\ast}e^{i\pi\sigma} -\chi^{\ast}e^{-i\pi\sigma}}
{\zeta e^{-i\pi\sigma} -\chi e^{i\pi\sigma}}
\epsilon ^{-2\rho},
\end{eqnarray}
where 
\begin{eqnarray}
  \zeta &=& \Gamma(2\rho)(2m^2M^2)^{-\rho}
e^{i\pi\rho},
\end{eqnarray}
and
\begin{eqnarray}
\chi &=&\Gamma(2\rho)(2m^2M^2)^{-\rho}
e^{-i\pi\rho}.
\end{eqnarray}
Substituting Eq.(\ref{exp_large-mass}) into Eq.(\ref{eq:branch-cut2}),
we obtain the contribution from Green's function to the asymptotic tail;
\begin{eqnarray}
\label{asympto-int_large-mass}
G^C(r_{\ast},r_{\ast}';t)
&\simeq&
\frac{1}{4\pi mi}
\tilde{\psi}_1(r_{\ast},m)
\tilde{\psi}_1(r'_{\ast},m)
\int _{\rm dashed\>line}
e^{2i\pi\sigma -2\rho \ln \epsilon -i\omega t} 
e^{i\varphi_{l}}
d\omega +({\rm complex \> conjugate }),
\end{eqnarray}
where the phase $\varphi_{l}$ is defined by
\begin{equation}
\label{phase_large}  
  e^{i\varphi _{l}}=\frac{\zeta ^{\ast}
+\chi ^{\ast}e^{-2i\pi \sigma }}
{\zeta +\chi e^{2i\pi \sigma }},
\end{equation}
and it remains in the range $0 \le\varphi_{l} \le 2\pi$, 
even if $\sigma $ becomes very large,
since we have
\begin{eqnarray}
  |\zeta|^2-|\chi|^2
&=&\Gamma(2\rho)\Gamma(-2\rho)\left(e^{2i\pi\rho}-e^{-2i\pi\rho}\right)
\>>\>0.
\end{eqnarray}
Because the terms $e^{-i\omega t}$, $e^{2i\pi\sigma}$ 
and $e^{-2\rho \ln \epsilon }$
in Eq.(\ref{asympto-int_large-mass}) 
are rapidly oscillating at very late times,
the saddle-point integration allows us to evaluate accurately 
the asymptotic behaviors.
Just like the small-mass case in the previous section,
using the parameter $a$ defined as Eq. (\ref{assume}),
the oscillation terms  $e^{i(2\pi\sigma -\omega t -2\rho \ln \epsilon )}$ in 
Eq.(\ref{asympto-int}) can be  rewritten into the form
\begin{eqnarray}
 \label{exp_imt_large}
e^{i(mt)^{1/3}f_{l}(a)}e^{-imt}
\end{eqnarray}
in the limits $mt \to \infty$ and $\omega /m \to 1$, 
by keeping $a$ to be finite. Here we have
\begin{eqnarray}
  f_{l}(a)\simeq \frac {\sqrt{2}\pi mM}{a} +a^2 
+\frac{4mM}{(mt)^{1/3}}\ln \left[\frac{2\sqrt{2}mMa}{(mt)^{1/3}}\right].
\end{eqnarray}
The saddle point can be approximated by
\begin{eqnarray}
  \label{a_0_large}
a_0 &\simeq& \left(\frac{\pi mM}{\sqrt{2}}\right)^{1/3} 
\end{eqnarray}
for asymptotically late times
\begin{eqnarray}
 \label{time_large}
 mt \gg mM .
\end{eqnarray}
Note that Eq. (\ref{a_0_large}) is identical with Eq. (\ref{a_0}).
Evaluating Eq. (\ref{asympto-int_large-mass}) 
using the saddle point integration, we obtain the asymptotic tail behavior
\begin{eqnarray}
    \label{eq:massivetail_large}
 G^{C}(r_{\ast},r'_{\ast};t) 
&\simeq & \frac{1}{2\sqrt 3}(2\pi)^{\frac 56}
(mM)^{\frac 13}(mt)^{-\frac 56} 
\nonumber\\
&&
\times
\sin(mt -3/2(2\pi mM)^{2/3}(mt)^{1/3}
-4 mM \ln \epsilon _0
- \varphi _{l}(\sigma _0)+3\pi/4 )
\nonumber\\
&&
\times \tilde{\psi}_1(r_{\ast},m)\tilde{\psi}_1(r_{\ast}',m).
\end{eqnarray}
The additional relation (\ref{time_large}) is necessary 
for the asymptotic tail (\ref{eq:massivetail_large})
to dominate.  
The oscillation has the  period $m/2\pi$ and is modulated by 
the two types of phase shifts.
First, the terms 
$(3/2)(2\pi mM)^{2/3}(mt)^{1/3}+4 mM \ln \epsilon _0  $ represent 
a monotonous phase shift.
Second, the term $\varphi _{l}(\sigma _0)$ represents a periodic phase shift
since it remains within $0 \le \varphi _{l}\le 2\pi$.
The decay rate of the 
asymptotic tail in Schwarzschild background
is $t^{-5/6}$, which 
remains  identical with that in nearly extreme Reissner-Nordstr\"{o}m 
background shown
in the paper I.

\subsection{Phase shift}
In the same manner with  the small mass case,
we study the periodic phase-shift effect
 which is caused by the term $e^{i\varphi _l}$ in
Eq.(\ref{phase_large}).
In the large mass limit $mM \gg 1$, 
the ratio of  $\chi $ and $\zeta$ is approximated by
\begin{eqnarray}
\label{ratio_large_S}
  \left|\frac{\chi}{\zeta}\right|&\simeq&e^{-\pi 4mM}\ll 1
\end{eqnarray}
in the Schwarzschild case.
On the other hand, in the nearly extreme Reissner-Nordstr\"{o}m limit,
we obtain the large mass limit from 
Eqs.(59) and (60) in paper I as follows,
\begin{eqnarray}
\label{ratio_large_R}
  \left|\frac{\chi}{\zeta}\right|_{\rm nearly \> extreme}&\simeq& e^{-\pi \sqrt{5}mM}\ll 1.
\end{eqnarray}
The oscillation amplitude $\Delta \varphi_{l}$ of $\varphi _l$ is 
\begin{eqnarray}
  \Delta \varphi _{l}&\simeq&  4 \left|\frac{\chi}{\zeta}\right|,
\end{eqnarray}
for which Eqs. (\ref{ratio_large_S}) and (\ref{ratio_large_R}) 
give different values,
though it remains same that $\Delta \varphi$ vanishes in the large-mass limit.

%%%%%%%%%%%%%%%%%%%%%%%%%%%%% Summary %%%%%%%%%%%%%%%%%%%%%%%%%%%%%%%%%%
\section{Summary}
In this paper we have investigated asymptotic tail behaviors
of massive scalar fields in Schwarzschild background.
We have shown that the asymptotic oscillatory
tail  with the decay rate of $t^{-5/6}$ 
and with the period $2\pi/m$
dominates, which is identical with that 
in nearly extreme Reissner-Nordstr\"{o}m background.
In addition, we have noted that the relation
 between the  value of $mM$ and
the time-scale when the $t^{-5/6}$ tail begins to dominate
also holds;
The smaller the value of $mM$ is, 
the later the $t^{-5/6}$ tail begins to dominate ($mt \gg 1/(mM)^2$).
The larger the value of $mM$ is, 
the later the $t^{-5/6}$ tail begins to dominate ($mt \gg mM$).

%%%%%%%%%%%

As far as the intermediate late-time behavior is concerned,
our result agrees with \cite{HandP}.  
However they claimed 
 "SI perturbation fields decay at late times slower than
any power law" in \cite{HandP}, 
which disagree with the expressions 
(\ref{eq:massivetail}) and  (\ref{asympto-int_large-mass}).
We believe that their simulation was not  carried out until
{\it enough} late times for the asymptotic tail to dominate 
or there occured numerical errors in their simulation.
Also, our result is consistent with \cite{Burko},
which was done numerically 
(both for a test field and for the fully nonlinear cases).

In terms of the phase shift
we can understand more clearly the tail behavior to be  a 
backscattering effect due to space-time curvature.
In the small-mass case $mM \ll 1$,
the intermediate tail with an  oscillatory power-law 
dominates
at the time scale $1\ll mt \ll 1/(mM)^2$
before the asymptotic tail discussed here
dominates \cite{KandT,HandP} (see also \cite{Burko}).
At both the  intermediate and asymptotic time scales
the frequency of a wave which contributes to tails is very close to  $m$.
However,  
the oscillation of the asymptotic tail is modulated  by 
phase shift effects 
which do not appear in the  intermediate tail. %\cite{KandT,HandP}.
There exist two types of the phase shift, which are
 monotonous and periodic with time.
The periodic phase shift crucially depends on the field mass.
The shift angle becomes nearly equal to 
  $2\pi $ in small mass limit $mM \to 0$,
and the shift oscillates  at the time scale  of $1/(m^3M^2)$,
which
is a long-term effect if
compared with the basic oscillation with the period of $2\pi /m$. 
In large mass limit $mM \to \infty$,
the shift angle become very small.
In scattering theory, it is well-known that 
the phase shift of wave is connected with
the presence of the scatterer.
In this case, of course, the scatterer is 
the effective potential of black-hole spacetime. 
Therefore it is clear that 
tails of massive scalar fields are  generated from  backscattering
%by curvature-induced potential.
due to space-time curvature.
The shift angle in Schwarzschild geometry 
is numerically different from that in nearly extreme
Reissner-Nordstr\"{o}m geometry,
but the qualitative behavior depending on $mM$ is similar.

We can conclude that the asymptotic tail with the qualitatively same behavior 
dominates 
both in Schwarzschild  and in nearly extreme
Reissner-Nordstr\"{o}m background.
Therefore it is conjectured that 
the oscillatory $t^{-5/6}$ tail 
caused by {\it resonance} backscattering
at asymptotic late times 
is a general feature
in {\it arbitrary} Reissner-Nordstr\"{o}m background.
It remains in  a future work to investigate 
what kind of instability of Cauchy horizons is caused by the massive tails. 

Finally we give a comment in comparison with the
effects induced by the field mass $m$ on vacuum polarization of
quantum massive scalar fields in the thermal state.
It has been found that the amplitude of vacuum polarization 
is enhanced around $mM\simeq O(1)$
 in the case of nearly extreme Reissner-Nordstr\"{o}m geometry,
while it is not seen in Schwarzschild case \cite{TandK}.
The field-mass induced effect on vacuum polarization
becomes clear only in the nearly extreme limit.
The reason will be explained as follows;
Vacuum polarization of massive scalar fields is due to both 
thermal excitation induced by black-hole temperature
and mass induced excitation.
In nearly extreme case, of which the black-hole temperature is nearly zero,
the mass induced effect becomes significant because   
the thermal excitation is suppressed.
On the other hand, in Schwarzschild case 
the mass induced effect is just hidden because 
the thermal effect by the black-hole temperature can  dominate.

We expect that resonance behavior due to the mass of a field 
interacting with a black hole may exist in various processes
as a generic feature of black-hole geometry. 
In order to confirm this conjecture,
it is necessary to investigate various processes 
of massive fields in more general black hole models.
This will be one of the interesting subjects in black-hole physics.

%%%%%%%%%%%%%%% acknowledgments %%%%%%%%%%%%%%%%%%%%%%%%%%%%%%%%%%%%%%%
\acknowledgments
The authors thank L. M. Burko and K. Nakamura for useful comments.

%%%%%%%%%%%%%%%%%%%%%%%% references %%%%%%%%%%%%%%%%%%%%%%%%%%%%

\end{document}